\documentclass[USenglish,twocolumn]{article}
\usepackage[utf8]{inputenc}				
\usepackage[big,online]{dgruyter}	
\usepackage{lmodern} 
\usepackage{microtype}
\usepackage{comment}
\usepackage[numbers,square,sort&compress]{natbib}
\usepackage{bm}
\usepackage{color} 
\theoremstyle{dgthm}

\theoremstyle{dgdef}

\begin{document}
  \articletype{Research Article}
  \received{Month	12, 2022}
  \revised{Month	DD, YYYY}
  \accepted{Month	DD, YYYY}
  \journalname{De~Gruyter~Journal}
  \journalyear{YYYY}
  \journalvolume{XX}
  \journalissue{X}
  \startpage{1}
  \aop
  \DOI{10.1515/sample-YYYY-XXXX}

\title{
  A symmetry-protected exceptional ring in a photonic crystal with negative index media}
\runningtitle{Short title}

\author*[1]{Takuma Isobe}
\author[2]{Tsuneya Yoshida}
\author[3]{Yasuhiro Hatsugai} 
\affil[1]{\protect\raggedright 
University of Tsukuba, Graduate School of Pure and Applied Sciences, Ibaraki, Japan, e-mail: isobe@rhodia.ph.tsukuba.ac.jp}
\affil[2]{\protect\raggedright 
Kyoto University, Department of Physics, Kyoto, Japan, email: yoshida.tsuneya.2z@kyoto-u.ac.jp}
\affil[3]{\protect\raggedright 
University of Tsukuba, Department of Physics, Tsukuba, Japan, e-mail: hatsugai@rhodia.ph.tsukuba.ac.jp}
	

\abstract{
Non-Hermitian topological band structures such as symmetry-protected exceptional rings (SPERs) can emerge for systems described by the generalized eigenvalue problem (GEVP) with Hermitian matrices.

In this paper, we numerically analyze a photonic crystal with negative index media, which is described by the GEVP with Hermitian matrices. Our analysis using COMSOL Multiphysics{\textregistered}  demonstrates that a SPER emerges for photonic crystals composed of split-ring resonators and metal-wire structures.
We expect that the above SPER can be observed in experiments as it emerges at a finite frequency.
}

\maketitle

\section{Introduction}
  The topological band theory has been studied as one of the central issues of condensed matter physics after the discovery of the integer quantum Hall effect~\cite{Klitzing_IQHE_PRL(1980),Laughlin_IQHE_PRB(1981),TKNN_IQHE_PRL(1982),Kohmoto_IQHE_AoP(1985),Haldane_AIQHE_PRL(1988)}.
Extensive studies have revealed the many types of topological phases and novel phenomena, such as bulk-boundary correspondence in electron systems~\cite{Hatsugai_BEC_PRL(1993),Hatsugai_BEC_PRB(1993)} or novel transport properties~\cite{W.Xiang_PRB_2011,Yang_PRB_2011,A.Birkov_PRL_2011,Xu_PRL_2011} on insulators~\cite{C.L.Kane_E.J.Mele_PRL.2005,C.L.Kane_E.J.Mele_PRL.2005_Z2,L.Fu_C.L.Kane_PRL.2007,M.Z.Hasan_C.L.Kane_RevModPhys.2010,X.L.Qi_S.C.Zhang_RevModPhys.2011,Y.Ando_JPSJ_2013,B.A.Bernevevig_T.L.Huglhes_S.C.Zhang_Science_2006,M.Knig_Science_2007,L.Fu_C.L.Kane_PRB.2007,L.Fu_C.L.Kane_PRB.2006,D.J.Thouless_PRB.1983,Schnyder_PRB.2008,A.Y.Kitaev_AIP_Conf_2009,S.Ryu_A.P.Schnyder_A.Furusaki_New.J.Phys_2010,X.L.Qi_T.L.Hughes_S.C.Zhang_PRB_2008,A.M.Essen_J.E.Moore_D.Vanderbilt_PRL_2009} or semimetals~\cite{S.Murakami_IOP_2007,W.Xiang_PRB_2011,Yang_PRB_2011,A.Birkov_PRL_2011,Xu_PRL_2011,Kurebayashi_JPSJ_2014,N.Armitage_RevModPhys_2018,Koshino_PRB_2016}.
Remarkably, classical systems are also within the scope of the application of the topological band theory: it has been studied in photonic systems~\cite{Raghu_PhC_PRL(2008),Raghu_PhC_PRA(2008),MIT_PhChIns_PRL(2008),Lu_TopPhot_Nat(2014),Hu_TopPhot_PRL(2015),Takahashi_Optica(2017),Takahashi_JPSJ(2018),Ozawa_TopPhot_RMP19,OtaIwamoto_NatPhoto(2020),Moritake_NanoPh(2021)}, phononic systems~\cite{Kariyado_MechGraph_Nat(2015),Yang_TopAco_PRL(2015),Huber_TopMech_Nat(2016),Susstrunk_MechClass_PNAS(2016),Tomoda_AIP(2017),Takahashi_Mech_PRB(2019),Liu_TopPhon_AFM(2020)}, electrical circuits~\cite{Lee_TopCir_Nat(2018)}, diffusion phenomena~\cite{Yoshida_Difus_Nat(2021),Makino_Difus_PRE(2022),Hu_ObsDifs_AM(2022)}, game theory~\cite{Knebel_GameTheor_PRL(2020),Yoshida_GameTheor_PRE(2021)}, and so on.
The ubiquity of the topological phenomena arises from the fact that the systems are described by the eigenvalue problem.

These studies for the quantum and the classical systems have recently been extended to non-Hermitian systems~\cite{Hatano_PRL(1996),Hatano_PRB(1998),K.Esaki_PRB_2011,M.Sato_Progress_of_Theoretical_Science_2012,S.-D.Liang_PRA_2013,D.Leykam_PRL_2017,Xu_PRL(2017)_WeylEP,Gong_class_PRX18,Yao_nHChern_PRL2019,Kawabata_nHclass_PRX19,Yoshida_nHFQH19,Xiao_ObsNHBBC_NatPhys(2020),xiao_ObsNBloch_PRL(2021),Dibyendu_PRB(2021)_nH-SSH,Bergholtz_nHer_RMP(2021)}.
This extension enriches the topological phenomena.
In particular, the non-Hermiticity induces unique phenomena, such as non-Hermitian skin effects~\cite{T.E.Lee_PRA_2016,Shunyu_PRL(2018)_SkinEffect,Flore_skin_PRL(2018),Yoshida_MSkinPRR20,Yokomizo_nbloch_PRL(2019),Borgina-Jan_PRL(2020),Okuma_skin_PRL(2020),CFang_skin_PRL(2020)} and the emergence of exceptional points (EPs)~\cite{Kozii_nH_arXiv(2017),Shen_NHTopBand_PRL(2018),Yoshida_NHhevferm_PRB(2018),Zyuzin_NHWyle_PRB(2018),Takata_pSSH_PRL(2018),Michishita_EPKondo_PRB(2020),Mandal_HighEP_PRL(2021),Delplace_EP3_PRL(2021)}.
The non-Hermitian skin effect is induced by point-gap topology, which results in extreme sensitivity to the presence/absence of the boundaries.
On the EPs, band touching occurs for both the real and the imaginary parts, which is also protected by point-gap topology.
The EPs are further enriched by symmetry [e.g., the interplay between pseudo-Hermiticity and EPs results in symmetry-protected exceptional rings (SPERs)~\cite{Zhen_ERing_nature(2015),Budich_SPERs_PRB(2019),Yoshida_SPERs_PRB(2019),Yoshida_SPERsMech_PRB(2019),Yoshida_PTEP(2020)}  (surfaces~\cite{Okugawa_SPERs_PRB(2019),Zhou_SPERs_Optica(2019),Kimura_SPES_PRB(2019)})  in two (three) dimensions].

Here, one may expect that further extension of the band theory results in novel topological phenomena. 
In addition, several systems (e.g., photonic and phononic systems) are described by the generalized eigenvalue problem (GEVP).
Recently, it has been reported that systems described by the GEVP with Hermitian matrices exhibit SPERs with emergent symmetry~\cite{IYH_PRB(2021)}. 
The emergence of such SPERs at the zero frequency explains the characteristic dispersion relation for hyperbolic metamaterials, which are continuum systems described by the Maxwell equations. 
However, studies of a lattice system hosting the above SPERs with the GEVPs is missing.

In this paper, we analyze a photonic crystal composed of negative index media (NIM), which is described by the GEVP with Hermitian matrices.
Our analysis using COMSOL Multiphysics{\textregistered} demonstrates that the above photonic system, composed of split-ring resonators (SRRs) and metal-wire structures, hosts a SPER with emergent symmetry at a finite frequency.
In this system, the negativity of the permittivity and the permeability results in the indefiniteness of the matrices, which is essential for the emergence of the above SPER.

The rest of this paper is organized as follows.
In Sec.~2, we briefly review GEVPs and analyze the band structure of a toy model. 
The toy model hosts a SPER protected by the emergent symmetry.
In Sec.~3, we analyze a photonic crystal composed of NIM.
We show the emergence of a SPER and discuss the observability of the SPER.
In Sec.~4, a short summary and discussion are provided.
In Appendix~A, Eq.~(2) is derived from Eq.~(1).
In Appendix~B, the relation between the pseudo-Hermiticity and SPERs is discussed.
In Appendix~C, a band structure for the positive permittivity and permeability is shown.
In Appendix~D, we analyze a $\omega$-dependent toy model by two different approaches.

%
\section{GEVP and non-Hermiticity}
In this section, we briefly review the GEVP and analyze a toy model~\cite{footnote3}.
In contrast to the ordinary eigenvalue problem, eigenvalues of GEVPs can be  complex numbers even when the matrices are Hermitian.
Whether the eigenvalues are real or complex is not determined by the Hermiticity of matrices in GEVPs, though it is determined only by the Hermiticity in the eigenvalue problems.
GEVP is defined by

\begin{equation}
H\psi = ES\psi,
\end{equation}
where $H$ and $S$ are Hermitian matrices. 
The eigenvalue is denoted by $E$, and $\psi$ is the eigenvector.

In this problem, the properties of eigenvalues significantly depend on whether the matrices are definite or indefinite.
In the case that $H$ or $S$ is definite (i.e., all eigenvalues have the same sign), all of eigenvalues are real.
On the other hand, in the case that the matrices are indefinite, eigenvalues can take complex values in spite of the Hermiticity.
This can be seen as follows.
Firstly, we note that Eq.~(1) can be rewritten as the following standard eigenvalue problem,
\begin{equation}
H_{\Sigma}\psi=E\psi.
\end{equation}
Here, $H_{\Sigma}$ is defined as $H_{\Sigma}=\Sigma H'$ with the Hermitian matrix $H'$ and the diagonal Hermitian matrix $\Sigma$ satisfying $\Sigma^2=1$ (for detailed derivation and explicit definitions of $H'$ and $\Sigma$, see Appendix A).  
We note that $\Sigma$ is the identity matrix unless $S$ is indefinite.
Namely, when $S$ is definite, Eq.~(1) is reduced to the ordinary eigenvalue problem [Eq.~(2)] with the Hermitian matrix, which results in $E\in \mathbb{R}$. When $S$ is indefinite, Eq.~(1) is reduced to the ordinary eigenvalue problem [Eq.~(2)]  with the non-Hermitian matrix~\cite{footnote1}, which may results in $E\in \mathbb{C}$.
In addition, $H_{\Sigma}$ satisfies a symmetry constraint
\begin{equation}
\Sigma H_{\Sigma}\Sigma=H_{\Sigma}^{\dagger}, 
\end{equation}
which is the pseudo-Hermiticity.
Hence, eigenvalues $E$ are real or form pairs $(E,E^*)$.
This pseudo-Hermiticity is attributed to the Hermiticity of the matrices $H$ and $S$.
From the above discussion, we can expect the emergence of the symmetry-protected non-Hermitian topology in the system described by the GEVP with indefinite Hermitian matrices.
\begin{figure}[t]
   \includegraphics[width=\hsize]{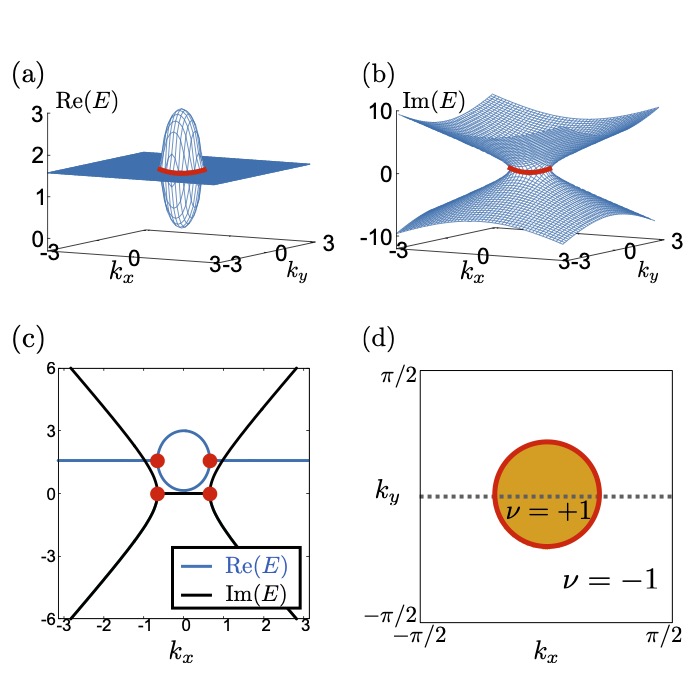}
 \caption{Band structure and $\mathbb{Z}_2$ invariant for $\bm{b}=(0, 0, 0.5)$ and $\bm{d}=(k_x, k_y, 0)$.
   Panel (a) [(b)] is the plot of the real [imaginary] part of the band structure. 
   Red lines indicate the SPER. 
   Panel (c) is the plot of the real (imaginary) part of the band structure at $k_y=0$.
   Blue (black) lines represent the real (imaginary) part. 
   Red points corresponds to the SPER [see also panel (d)]. 
   Panel~(d) represent the $\mathbb{Z}_2$ invariant $\nu$.
   Orange (white) region is $\nu=+1$ ($\nu=-1$).
   Red ring corresponds to the SPER in panels~(a) and (b).
   Band structures in panel~(c) are ploted along the dotted line.}
 \label{fig:fig1}
\end{figure}

In order to show the emergence of the non-Hermitian band structures with Hermitian matrices, we analyze a two-band model described by the GEVP, whose matrices are given by
\begin{align*}
  H(\bm{k})&=\sum_{i=0}^3 h_i(\bm{k})\sigma_i, & S(\bm{k})&=\sum_{i=0}^3 s_i(\bm{k})\sigma_i.
\end{align*}
Here, $h_i(\bm{k})$ and $s_i(\bm{k})$ are real, and $\sigma_0$ is the $2\times2$ identity matrix, $\sigma_i$ are the Pauli matrices ($i=1,2,3$). 
The vector $\bm{k}=(k_x,k_y)$ describes the momentum.
Its eigenvalues are given by
\begin{equation}
  E(\bm{k})_{\pm}=\frac{1}{\eta(\bm{s},\bm{s})}[\eta({\bm{h},\bm{s}})\pm\sqrt{\eta(\bm{h},\bm{s})^2-\eta(\bm{h},\bm{h})\eta(\bm{s},\bm{s})}],
\end{equation}
where $\eta( , )$ is the Minkouski product, $\eta(\bm{a},\bm{b})=a_0b_0-a_1b_1-a_2b_2-a_3b_3$~\cite{footnote2}.

Here, we analyze the model with $\bm{h}(\bm{k})=(0,k_x,k_y,M_L)$, $\bm{s}(\bm{k})=(1,0,0,M_R)$.
We choose as $|M_L|>0$, $|M_R|>1$, so that $H$ and $S$ are indefinite.
Figure~1 is the dispersion relations with  $M_L=0.3$ and $M_R=1.1$.
Figure~1(a) [1(b)] shows the real [imaginary] part of the band structure.
One can see emergence of the symmetry-protected exceptional ring, which is represented as the red line.
In Fig.~1(c), the real and the imaginary parts of eigenvalues at $k_y=0$ are shown.
The red dots denote EPs.
At the red dots, the band touching occurs both for the real and the imaginary parts. 

The above ring of the exceptional points is protected by the pseudo-Hermiticity. 
For our $2\times 2$ model, the GEVP is rewritten as $H_{\Sigma}\psi=E\psi$, with
\begin{equation}
H_{\Sigma}=
  \begin{pmatrix}
    \frac{M_L}{|1+M_R|}&\frac{k_x-ik_y}{\sqrt{|1-M_R^2|}}\\
    -\frac{k_x+ik_y}{\sqrt{|1-M_R^2|}}& \frac{M_L}{|1-M_R|}
  \end{pmatrix}.
\end{equation}
This matrix possesses pseudo-Hermiticity for the operator $\Sigma=\sigma_3$, $\sigma_3H_{\Sigma}\sigma_3=H_{\Sigma}^{\dagger}$.

The presence of the pseudo-Hermiticity allows us to define the following $\mathbb{Z}_2$-invariant,
\begin{equation}
\nu =\mathrm{sgn} \Delta(\bm{k}).
\end{equation}
Here, $\Delta(\bm{k})$ is the discriminant of the polynomial of $E$, $\mathrm{det}[H(\bm{k})-ES(\bm{k})]= \mathrm{det}[S(\bm{k})]\mathrm{det}[H_{\Sigma}(\bm{k})-E]  =a_N(-E)^N+a_{N-1}(-E)^{N-1}+\cdots +a_1(-E)+a_0$ with $a_i \in \mathbb{C}$. 
It is defined as 
\begin{equation}
  \Delta(\bm{k})=\prod_{n<n'}[E_n(\bm{k}) - E_{n'}(\bm{k})]^2,
\end{equation}
where $n$ label eigenvalues $E_n$ ($n=1,\ldots,N$).
Here two remarks are in order. 
(i) Because of the pseudo-Hermiticity, $\Delta$ is real.
We recall that $\mathrm{det}S(\bm{k})$ does not change its sign in the momentum space.
(ii) The discriminant can be computed only from the coefficients $a_i$. 
In Fig.~1(d), the $\mathbb{Z}_2$-invariant $\nu$ for each point in the momentum space is shown. 
The SPER emerges on the line where $\nu$ jumps because at least two roots $E_n$ are equal for $\Delta =0$.

Here we have shown that the system described by the GEVP with Hermitian matrices exhibits the SPER protected by the emergent symmetry [see Eq.~(3)] when both matrices are indefinite [see Eq.~(1)].
In the next section, we analyze a photonic crystal composed of NIM described by the GEVP where the above SPER emerges at a finite frequency.
\begin{figure}[t]
   \includegraphics[width=\hsize]{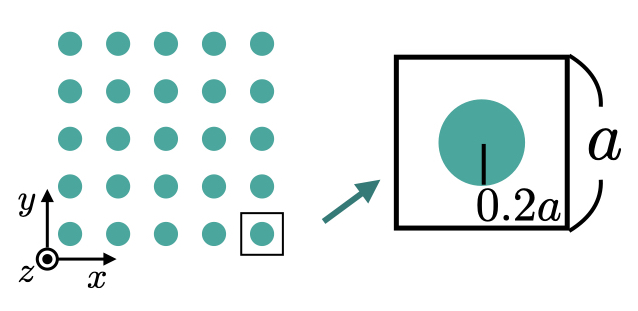}
 \caption{
   Sketch of the photonic crystal. 
   Unit-cell size is represented $a$ and the radius of internal structures of the photonic crystal is chosen as $0.2a$.
The regions of NIM (vacuum) are colored in green (white).
The permittivity and the permeability are chosen in $\varepsilon=-5.9$ and $\mu=-0.4$ ($\varepsilon=\mu=1$) in the green- (white-) colored region.
 }
 \label{fig:fig1}
\end{figure}

\begin{figure*}[t]
   \includegraphics[width=\hsize]{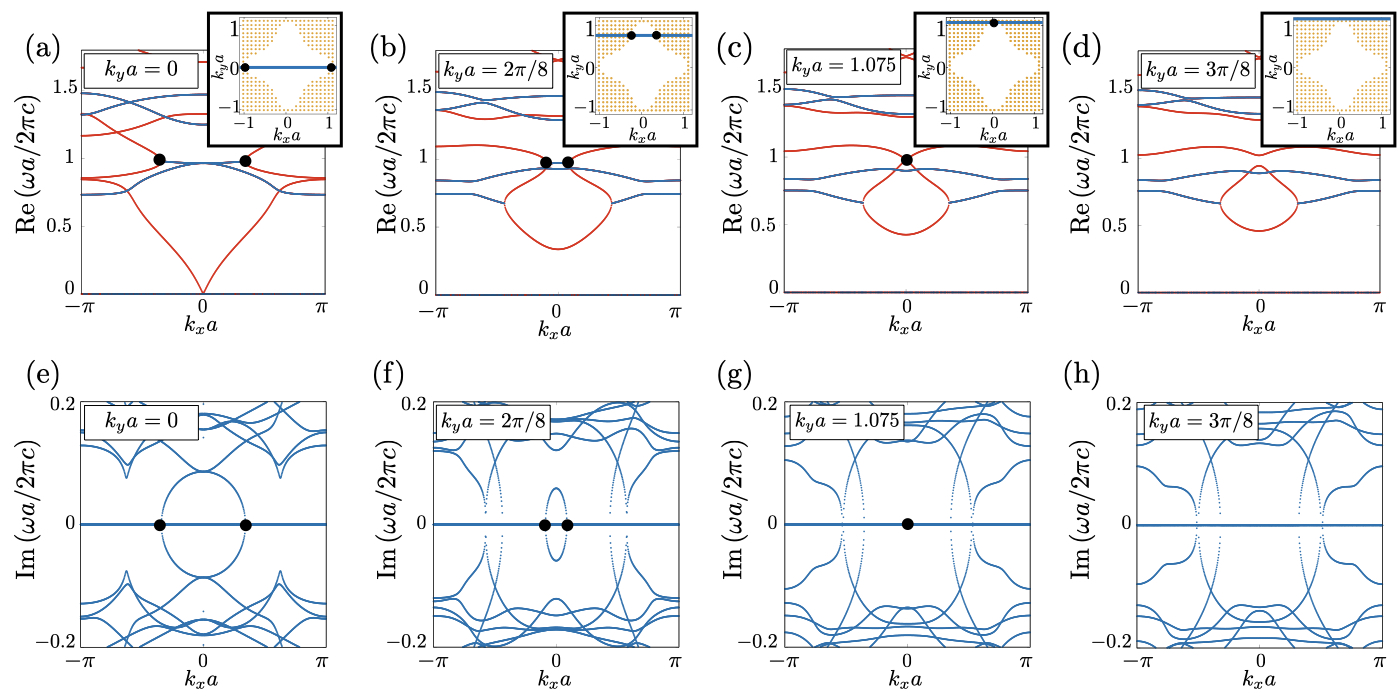}
 \caption{
Photonic band structures of the square lattice photonic crystal composed of NIM.
The relative permittivity and the relative permeability are fixed at $\varepsilon=-5.9$ and $\mu=-0.4$ [see Fig.~2].
The radius of internal structures of the photonic crystal is chosen as $0.2a$, as illustrated in Fig.~2.
Panels (a)-(d) are the plot of the real part for several values of $k_ya$ denoted by the blue lines in the insets. 
The bands of real eigenvalues are colored in red. 
The SPERs are represented by black dots. 
In the insets of these panels, the $\mathbb{Z}_2$-invariant $\nu$ is plotted; $\nu$ takes $1$ and $-1$ in the orange- (white-) colored regions. 
The boundary between these regions corresponds to the SPER on which we focus. The section between SPER and a line of specific $k_ya$ is represented by black dots.
 }
\end{figure*}


\section{Photonic crystal with negative index media}
\subsection{Photonic band calculation and SPER}

In this section, we analyze a photonic crystal composed of NIM, which hosts  a SPER at a finite frequency.
For the emergence of the non-Hermiticity based on the GEVP, systems need to satisfy the Hermiticity and the indefiniteness of the matrices.
Photonic systems are the one of the systems which can satisfy the above properties.

The band structure of photonic crystals are described by the Maxwell equations,
\begin{align*}
  \bm{\nabla}\times\bm{E}&=-\frac{\partial \bm{B}}{\partial t},
  &\bm{\nabla}\times\bm{H}&=\frac{\partial \bm{D}}{\partial t},\\
  \bm{\nabla}\cdot\bm{D}&=0,
  &\bm{\nabla}\cdot\bm{B}=0.
\end{align*}
Here, $\bm{E}$ ($\bm{H}$) is the electric (magnetic) field, and $\bm{D}$ ($\bm{B}$) is the electric (magnetic) flux density.
From these equations, we obtain the following wave equation,
\begin{equation}\label{wave}
  \bm{\nabla}\times\mu(\bm{r})^{-1}\bm{\nabla}\times\bm{E}=\left(\frac{\omega}{c}\right)^2\varepsilon(\bm{r})\bm{E},
\end{equation}
 where $\varepsilon$ and $\mu$ are the relative permittivity and the relative permeability.
 Angular frequency, position in real space, and light speed in vacuum are represented as $\omega$, $\bm{r}$, and $c$, respectively.

The above equation is a GEVP. 
To see this, we expand the electric field $\bm{E}(\bm{r})$ by using real space bases as $\bm{E}(\bm{r})=\sum_j |\bm{\phi}_j\rangle c_j$. 
Here, $|\bm{\phi}_j\rangle$ is the localized bases on the mesh point $j$. 
Applying $\langle \bm{\phi}_i |$ from the left yields,
  \begin{align}\label{GEVPMaxwell}
  \sum_j\langle\bm{\nabla}\times\bm{\phi}_i|&\mu^{-1}(\bm{r})|\bm{\nabla}\times\bm{\phi}_j\rangle c_j \notag\\ 
  &=\sum_j\left(\frac{\omega}{c}\right)^2\langle\bm{\phi}_i|\varepsilon(\bm{r})|\bm{\phi}_j\rangle c_j.
\end{align}
The above equation is a GEVP which can be seen by noting the correspondence, $H_{ij}\leftrightarrow \langle \bm{\nabla}\times\bm{\phi}_i | \mu^{-1}(\bm{r}) |\bm{\nabla}\times\bm{\phi}_j \rangle $, $S_{ij}\leftrightarrow \langle\bm{\phi}_i|\varepsilon(\bm{r})|\bm{\phi}_j\rangle$, $\bm{\psi}_j \leftrightarrow c_j$.
Here, one might consider that the minus sign of $\mu$ and $\varepsilon$ are factored out by focusing only on the region where both $\mu$ and $\varepsilon$ are negative. However, such a procedure is not applicable because electromagnetic fields can propagate to the outside of regions where $\mu$ and $\varepsilon$ are negative; electromagnetic fields are not confined in the region where $\mu$ and $\varepsilon$ are negative.
Indeed, choosing $\mu$ and $\varepsilon$ to positive values significantly changes the band structure (for more details, see Appendix C).

As discussed in Sec.~2, breaking the definiteness of the both matrices $H$ and $S$ is essential for the emergence of non-Hermitian band structures.
 Breaking the definiteness can be accomplished by negative $\varepsilon$ and $\mu$ in the internal structures of the photonic crystal.

Now, we consider the photonic crystal of the square lattice~[see Fig.~2].
Here, we take the $x$-$y$ plane parallel to the two-dimensional photonic crystal and the $z$-axis perpendicular to the system. 
The unit-cell size is denoted by $a$, and the radius of the internal structures of the photonic crystal is set to be $0.2a$.
Here, we consider the case when $\varepsilon$ and $\mu$ respectively take nagative constant values in the green-calored region of Fig.~2.
The permittibity and the permeability of green- (white-) colored region are chosen in $\varepsilon=-5.9$ and $\mu=-0.4$ ( $\varepsilon=1$ and $\mu=1$ ).
We assume the photonic crystal is uniform and infinitely long in the $z$-direction.

Figure~3 shows the photonic band structures of the transverse magnetic (TM) mode [$E=(0,0,E_z)$, $H=(H_{x},H_{y},0)$].
Eigenvalues are calculated for each $k$ by assuming $\varepsilon$ and $\mu$ are constant.
The data are obtained by using COMSOL. 
Specifically, we employ the wave optics module. 
We set ``physics controlled mesh" to fine mesh size.
In Figs.~3(a)-3(d) [3(e)-3(h)], the real [imaginary] part of the dimensionless parameter $\omega a/(2\pi c)$ are plotted for  several values of $k_y a$ [see also the insets of Figs.~3(a)-3(d)].
The bands of real eigenvalues are plotted in red.
As denoted by the black dots, band touching occurs for both of the real and the imaginary parts, indicating emergence of the EPs at fixed $k_ya$ [see Figs.~3(a)-3(c) and 3(e)-3(g)]. 
These data indicate the emergence of the SPER in the two-dimensional momentum space~\cite{footnote5,footnote6}. 
We note that EPs are not observed in Figs.~3(d)~and~3(h) because the SPER does not cross the line specified by $k_ya=3\pi/8$ [see the inset of Fig.~3(d)].

Now, we address the topological characterization of the SPER by the $\mathbb{Z}_2$-invariant.
For the computation of $\nu$, we pick up the two bands involved in the SPER. The $\mathbb{Z}_2$-invariant is plotted in the inset of Fig.~3(b) [The inset of Figs.~3(a),~3(c)~and~3(d) show the same data of $\nu$]. The inset of Fig.~3(b) indicates that the SPER is indeed characterized by the $\mathbb{Z}_2$-invariant; the $\mathbb{Z}_2$-invariant $\nu$ jumps from $-1$ to $1$ on the SPER with increasing $k_xa$ from $0$ to $\pi$ [see Fig.~3(b)].

From the above results of the band structure and the $\mathbb{Z}_2$-invariant, we conclude that the photonic crystal composed of NIM hosts the SPER protected by emergent symmetry.

Here, a comment is in order concerning the physical meaning of the complex band structure. 
Our system does not include any dissipation, although the eigenvalues become complex.
This result may suggest that when shining electromagnetic fields onto our system, it is expected that the electromagnetic fields will spatially decay, as is the case of electromagnetic fields in photonic band gaps or plasmon in metals. More detailed discussion is conducted in Sec.~3.2.
Our SPER is defined as the boundary of the momentum space where propagating modes are present and absent.

\subsection{Photonic crystal composed of SRRs and metal-wire structures}

In Sec.~3.1, we have analyzed the photonic crystal composed of the negative $\varepsilon$ and $\mu$.
Although we have assumed $\varepsilon$ and $\mu$ are negative constant in the internal structures [i.e., the green-colored region of Fig.~2], $\varepsilon$ and $\mu$ are the function of the frequency for generic NIM. 
Here, let us discuss the case of the green colored region of Fig.~2 is composed of metal-wire structures~\cite{Pendry_MWS_JOP(1998)} and SRRs~\cite{Pendry_SRR_IEEE(1999)} [see Fig.~4(a)], which negative $\varepsilon$ and negative $\mu$ are experimentally reported~\cite{Smith_NIM_PRL(2000),Smith_NIM_Science(2001)}.
With the effective medium approximation, the permittivity and the permeability of NIM composed of SRRs and metal-wire structures are given by~\cite{Smith_NIM_Science(2001)},
\begin{equation}
  \frac{\varepsilon (\omega)}{\varepsilon_0} = 1-\frac{\omega_{\mathrm{ep}}^2-\omega_{\mathrm{eo}}^2}{\omega^2-\omega_{\mathrm{eo}}^2+i\gamma\omega},
\end{equation}
\begin{equation}
  \frac{\mu (\omega)}{\mu_0} = 1-\frac{\omega_{\mathrm{mp}}^2-\omega_{\mathrm{mo}}^2}{\omega^2-\omega_{\mathrm{mo}}^2+i\gamma\omega},
\end{equation}
where $\omega_{\mathrm{ep}}/2\pi=12.8~\mathrm{[GHz]}$, $\omega_{\mathrm{eo}}/2\pi=10.3~\mathrm{[GHz]}$, $\omega_{\mathrm{mp}}/2\pi=10.95~\mathrm{[GHz]}$, $\omega_{\mathrm{mo}}/2\pi=10.05~\mathrm{[GHz]}$, and $\gamma=10~\mathrm{[MHz]}$.
$\varepsilon_0$ and $\mu_0$ are the permittivity and the permeability in the vacuum, $\varepsilon_0=8.85\times 10^{-12}~[\mathrm{F/m}]$ and $\mu_0=1.25\times 10^{-6}~[\mathrm{H/m}]$.
The frequency dependence of $\varepsilon$ and $\mu$ is shown in Fig.~4(b). 
Both $\varepsilon$ and $\mu$ take negative values within the colored region in Fig.~4(b).
When $\omega/2\pi=10.7[\mathrm{GHz}]$, the permittivity and the permeability become $\varepsilon=-5.9$ and $\mu=-0.4$.
We note that $\gamma=10~[\mathrm{MHz}]$ is negligible compared to $\omega_{\mathrm{ep}}$, 
$\omega_{\mathrm{eo}}$, 
$\omega_{\mathrm{mp}}$, 
$\omega_{\mathrm{mo}}$ and 
$\omega$.
Therefore, Eq.~(9) corresponds to solving the following equation,
\begin{align}\label{GEVPMaxwell}
  \sum_j\langle\bm{\nabla}\times\bm{\phi}_i|&\mu^{-1}(\omega_c,\bm{r})|\bm{\nabla}\times\bm{\phi}_j\rangle c_j \notag\\ 
  &=\sum_j\left(\frac{\omega}{c}\right)^2\langle\bm{\phi}_i|\varepsilon(\omega_c,\bm{r})|\bm{\phi}_j\rangle c_j.
\end{align}
with $\omega_c/2\pi=10.7[\mathrm{GHz}]$.
Since $\omega a/2 \pi c$ is plotted as the vertical axis in Fig.~3, we can regard the vertical axis as the unit-cell size $a$ by fixing $\omega$ in $\omega_c$, and considering the vertical axis as $\omega_c a/2 \pi c$.

From the above perspective, Fig.~3 represents the ``band structure" of the unit-cell size $a$ that hosts the eigenmodes with real $\omega$ and real $k$.We recall that $\omega$ and $k$ are fixed in real in our analysis.
In the complex region of the band structure, $a$ needs to be complex for real $\omega$ and real $k$, although complex $a$ cannot be realized physically.
Therefore, the complex region of the band structure represents the area where the eigenmodes with real $\omega$ and $k$ cannot be excited physically.
This result indicates that as we vary $k$ along the band structure, physically excitable eigenmodes vanish at specific $k$ points. These $k$ points correspond to EPs, and the EPs surround the region where physically excitable eigenmodes are absent with real $\omega$ and real $k$.
Such absence of bands is unique to the GEVP composed of indefinite Hermitian matrices.
Here, we note that the spatially decaying modes with complex $k$ are not forbidden in the complex region~\cite{footnote9}.
Therefore, electromagnetic fields are considered to decay spatially as is the case of the electromagnetic fields in photonic band gaps or plasmon in metals.


We finish this section by the discussion towards the experimental observation. 
In the above, we have fixed the frequency. 
Thus, we consider that the square-root dispersion can be observed by changing the unit-cell size~\cite{footnote8}.
The square-root dispersion for a photonic crystal with dissipation is experimentally observed in Ref.~\cite{Zhen_ERing_nature(2015)}.  
Specifically, we expect that  the SPER is observed for approximately $a=2.8~[\mathrm{cm}]$ and $\omega/2\pi=10.7~[\mathrm{GHz}]$  because the SPER emerges at the vicinity of $\mathrm{Re}(\omega a/2\pi c)=1$.
The radius of the internal structures of the photonic crystal is approximately $0.2a=0.56~[\mathrm{cm}]$, which is larger than the size of SRRs, $0.5~[\mathrm{cm}]$, of Ref.~\cite{Smith_NIM_Science(2001)}.
We note that other metamaterials~\cite{Xu_NIM_Nature(2013)} can also be available as NIM while we have focused on the systems composed of SRRs and metal-wire structures.

\begin{figure}[t]
   \includegraphics[width=\hsize]{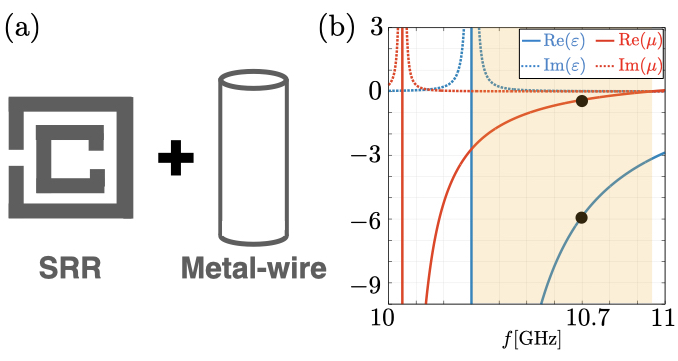}
 \caption{
Sketch of the SRR and the metal-wire structure, and the permittivity and permeability.
Panel (a) displays the sketch of the SRR and the metal-wire structure as a NIM.
Panel (b) displays the plot of the permittivity and the permeability of the NIM composed of SRRs and metal-wire structures. We fix the frequency to $\omega/2\pi=10.7~[\mathrm{GHz}]$, as denoted by black points.
}
\end{figure}

\section{Summary and Discussion}

In this paper, we have analyzed the photonic crystal composed of NIM, which is described by the GEVP with Hermitian matrices. 
By using COMSOL, we have elucidated that the photonic crystal exhibits the SPER protected by the emergent symmetry at a finite frequency. 
Here, the negativity of both the permittivity $\varepsilon$ and the permeability $\mu$ is essential for the emergence of the SPER  because it results in the indefiniteness of the relevant matrices.

So far, we have solved the GEVP by fixing $\omega$. 
In this case, the unit-cell size $a$ corresponds to eigenvalues.
Namely, the band structure of real eigenvalues in Fig.~3 can be observed by changing $a$ for fixed frequency $\omega$. 
We note that solving $\omega$ self-consistently in Eqs.~(9)-(11) also yields the band structure of real $\omega$ for our system. 
In contrast to our band structure, this band structure of self-consistent approach can be observed by inserting light with various values of frequency $\omega$ for fixed $a$. 
Although comparison between these two different band structures is not straightforward, these two band structures for a specific $\omega$ are consistent for the fixed $\omega$. Indeed we confirm the consistency for a toy model (for more details, see Appendix D).

In our system, complex eigenvalues emerge without dissipation.
Our results imply that as is the case of electromagnetic fields in photonic band gaps or plasmon in metal, the electromagnetic fields in the photonic crystal of the NIM would decay spatially in the region of the momentum space where bands take complex values.

In Ref.~\cite{IYH_PRB(2021)}, it has pointed out that the hyperbolic metamaterials host SPERs at zero frequency. 
In contrast to the previous work, we have demonstrated that our photonic crystal hosts the SPER at a finite frequency.
This SPER at a finite frequency is expected to be observed by changing the size of the unit-cell around  $a=2.8~\mathrm{[cm]}$.
We expect that the emergence of SPER will result in a unique reflection spectrum. The detailed analysis of the reflection spectrum and the time-evolution are left as future works.

We have employed the effective medium approximation in order to take into account the internal structure of the photonic crystal.
This approximation become accurate when the number of resonators inside of the cylinder is large. 
This fact indicates that our analysis becomes accurate for SPERs with large $\omega a/2\pi c$. 
We have also assumed that the system extends infinitely in the $z$-direction.  
For more accurate band structures, we need to take into account effects of the detailed internal structure of cylinders~\cite{footnote7} and effects of the boundaries in the $z$-direction. We left further analysis as a future work.

We note that the SPER attracts interests due to its application to high-sensitivity sensors~\cite{Wiersig_EPsensor_optica(2020)}. Our results of phtonic crystal might serve as a platform of such a novel device. 
We also note that if we scale down our system to nano-scale the frequency is around the visible light or nearinfrared regions.

\section{ACKNOWLEDGEMENT}
We thank Satoshi Iwamoto for fruitful discussions.
This work is supported by MEXT-JSPS Grant-in-Aid for Transformative Research Areas (A) ``Extreme Universe": Grant No.~JP22H05247. This work is also supported by JST-CREST Grant No.~JPMJCR19T1, JST-SPRING Grant No.~JPMJSP2124, and JSPS KAKENHI Grant No.~JP21K13850.


\begin{thebibliography}{100}
\bibitem{footnote1}
{ We note that in general $(\Sigma H')^{\dagger} \neq \Sigma H'$ because
  $\Sigma$ is not the identity matrix for indefinite $S$ }.
\bibitem{footnote2}
{ The cahracteristic equation of the $2\times 2$ GEVP is\\
  $\mathrm{det}(H-ES)=(h_0-Es_0)^2-\sum_{i=1}^3(h_i-Es_i)^2\\
  =(h_0^2-\sum_{i=1}^3h_i^2)+E^2(s_0^2-\sum_{i=1}^3s_i^2)\\-2E(h_0s_0-\sum_{i=1}^3h_is_i)\\
  =E^2\eta(\bm{s},\bm{s})-2E\eta(\bm{h},\bm{s})+\eta(\bm{h},\bm{h})$.\\ We can
  obtain Eq.~(4) by solving the quadratic equation
  $E^2\eta(\bm{s},\bm{s})-2E\eta(\bm{h},\bm{s})+\eta(\bm{h},\bm{h})=0$ }.
\bibitem{footnote5}
{The emergence of SPERs is also understood in terms of competition between the
  Hermitian part and the anti-Hermitian part of the matrix. Non-Hermitian
  matrices can be described by the summation of a Hermitian matrix and an
  anti-Hermitian matrix. Discarding the anti-Hermitian matrix, we can obtain
  the band structure with real values described by the Hermitian matrix.
  Introducing the anti-Hermitian matrix, the small band gap becomes pure
  imaginary, which results in the emergence of SPER.}
\bibitem{footnote9}
{Complex $\omega$ results in the complex $\varepsilon$ and $\mu$ from Eq.~(10)
  and Eq.~(11). We fix $\omega$ in real since complex $\varepsilon$ ($\mu$)
  breaks the Hermiticity of the matrices composing the GEVP.}
\bibitem{footnote8}
{Based on the band structures in Fig. 3, the unit-cell size $a$ and the wave
  number $k_x$ are determined. Firstly, we select a point on the band structure
  and determine the unit-cell size $a$ by referencing the vertical axis of the
  band structure since $\omega /2\pi c$ is constant. Once $a$ is determined,
  $k_x$ can be determined by referencing the horizontal axis of the band
  structure. The above discussion is based on changing $a$ while fixing
  $\omega$.}
\bibitem{footnote7}
{We consider that the band structure is not significantly affected at a
  specific wavenumber where the electromagnetic field become small inside of
  cylinders.}
\end{thebibliography}



\section*{APPENDIX A: GEVP and non-Hermiticity}
Here, let us derive Eq.~(2) from Eq.~(1).
First, we diagonalize matrix $S$ by a unitary matrix $U$, $U^{-1}SU=S_U$.
Correspondingly, $H$ and $\psi$ are also transformed $U^{-1}HU= H_U$, and $U^{-1}\psi = \psi_U$.

Second, we decompose matrix $S_U$ as follows,
\begin{equation}
S_U=S^{\prime}\Sigma S^{\prime}.
\end{equation}
  Here, $S^{\prime}=\mathrm{diag}(|\sqrt{s_1}|,\cdots,|\sqrt{s_N}|)$, $s_i$ are diagonal components of $S_U$. $\Sigma$ is a diagonal matrix whose diagonal elements are $+1$ or $-1$.
Note that when the matrix $S$ is definite, the matrix $\Sigma$ is equal to the identity matrix.

Finally, we set
\begin{align}
  H^{\prime}&=S^{{\prime}{-1}}H_US^{{\prime}{-1}},\\
  \psi^{\prime}&=S^{{\prime}{-1}}\psi_U,
\end{align}
  and operate $\Sigma$ from the left side. 
  We can obtain the eigenvalue problem
\begin{equation}
  H_{\Sigma}\psi^{\prime}=E\psi^{\prime},
\end{equation}
with $H_{\Sigma}=\Sigma H^{\prime}$.
The matrix $H^{\prime}$ is Hermitian because the transformation from $H$ to $H^{\prime}$ preserve the Hermiticity of the matrix $H$.

Therefore, in the case that $S$ is definite, $H_\Sigma$ becomes the Hermitian matrix. 
In contrast, if $S$ is indefinite, $H_\Sigma$ becomes non-Hermitian. 

\section*{APPENDIX B: Pseudo-Hermiticity and SPER}
In this section, we discuss the relation between the pseudo-Hermiticity and the symmetry-protected exceptional rings (SPERs), which is a symmetry-protected non-Hermitian topological band structure in a two-dimensional system~\cite{Zhen_ERing_nature(2015),Budich_SPERs_PRB(2019),Yoshida_SPERs_PRB(2019),Yoshida_PTEP(2020),Okugawa_SPERs_PRB(2019),Zhou_SPERs_Optica(2019)}.

We consider a two-band model of the non-Hermitian system, which Hamiltonian is given by
\begin{equation}
  H(\bm{k})=\sum_{i=1}^3\left[b_i(\bm{k})+id_i(\bm{k})\right]\sigma_i,
\end{equation}
where $b(\bm{k})$ and $d(\bm{k})$ are real, and $\sigma_i$ are Pauli matrices. 
We assume $\bm{k}$ is two-dimentional parameter.
$b_i$ represents the Hermitian part of the Hamiltonian. Non-Hermiticity of the Hamiltonian comes from $d_i$.
In general, $b_i$ and $d_i$ are three-dimentional vectors which directions and norms vary independently.
Eigenvalues are given by
\begin{equation}
E(\bm{k})_{\pm}=\pm\sqrt{b(\bm{k})^2-d(\bm{k})^2+2i\bm{b}(\bm{k})\cdot\bm{d}(\bm{k})}.
\end{equation}
EPs emerges at $\bm{k}$ points that satisfy $b^2-d^2=0$ and $\bm{b}\cdot\bm{d}=0$.
Figure~5(a) [5(b)] shows the real [imaginary] part of the band structure with $\bm{b}=(0, 0.5, 0.5)$ and $\bm{d}=(k_x, k_y, 0)$.
Red points indicate EPs.

Here, let us consider the case of the pseudo-Hermitian Hamiltonian.
The pseudo-Hermitian operator is chosen by $\Sigma=\sigma_3$, the third component of the Pauli matrices.
In this case, Hamiltonian with $\bm{b}=(0, 0, b_3)$ and $\bm{d}=(d_1, d_2, 0)$ preserve the pseudo-Hermiticity.
When the Hamiltonian has the pseudo-Hermiticity, one of the constraints for the band touching, $\bm{b}\cdot\bm{d}=0$, is satisfied automatically. 
Therefore, EPs form one-dimensional lines since one of the band touching conditions in the two-dimensional parameter space is eliminated. 
This structure is the SPER, which is one of the symmetry-protected non-Hermitian topological band structures.
Figures~5(c) and 5(d) are the band structure with $\bm{b}=(0, 0, 0.5)$ and $\bm{d}=(k_x, k_y, 0)$.
We can see the emergence of a SPER denoted by red lines.

\begin{figure}[t]
   \includegraphics[width=\hsize]{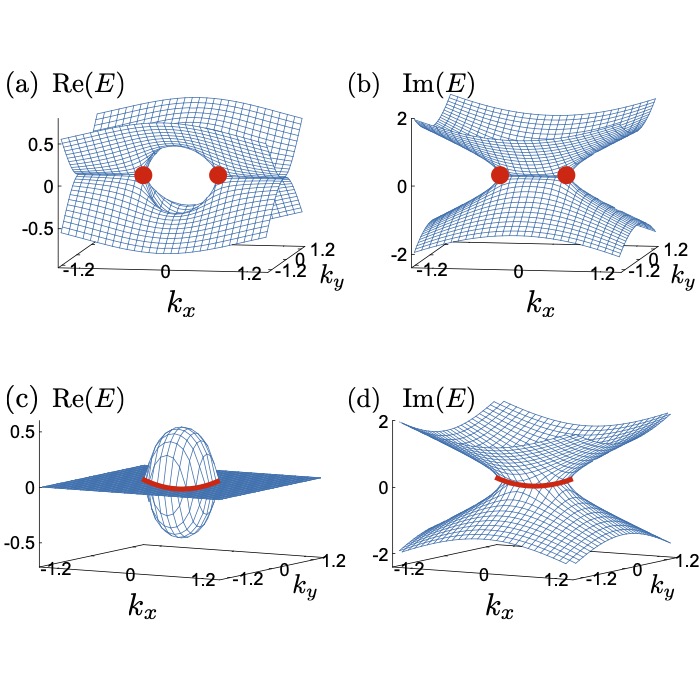}
 \caption{
   Panel (a) [(b)] is the plot of the real [imaginary] part of the band structure with $\bm{b}=(0, 0.5, 0.5)$ and $\bm{d}=(k_x, k_y, 0)$. Red points indicate EPs. Panel (c) [(d)] is the plot of the real [imaginary] part of the band structure with $\bm{b}=(0, 0, 0.5)$ and $\bm{d}=(k_x, k_y, 0)$. Red lines indicate the SPER.   
 }
 \label{fig:fig1}
\end{figure}

\section*{APPENDIX C: Band structure for positive permittivity and permeability}
We show that the sign of $\mu$ and $\varepsilon$ significantly affects on the band structure by comparing the band structures for positive $(\varepsilon,\mu)$ and negative $(\varepsilon,\mu)$.

Figure~6 provides the results of positive $(\varepsilon, \mu)$. 
As shown in this figure, the band structure is significantly different from the one shown in Fig.~3(a)~and~3(e). 
In particular, EPs do not emerge for $(\varepsilon, \mu)=(5.9,0.4)$ because the frequency $\omega$ remains real in the entire region.

The above difference of the band structure [see Fig.~3(a)~and Fig.~6] can be understood by analyzing a simple model.
Consider a system where electromagnetic fields can be expanded as 
$\bm{E}(\bm{r})=|\bm{\phi}_1\rangle \bm{c}_1+ |\bm{\phi}_2\rangle \bm{c}_2$.
Here, the $|\bm{\phi}_1\rangle$ ($|\bm{\phi}_2\rangle$) is bases located on th region 1 (2) where the permittivity and permeability are $\varepsilon_1$ and $\mu_1$ ($\varepsilon_2$ and $\mu_2$).
We suppose that the overlap between $|\bm{\phi}_1\rangle$ and $|\bm{\phi}_2\rangle$ is finite, which is reasonable for photonic crystals.
In this case, Eq.~(9) is rewritten as
\begin{equation}
\begin{pmatrix}
h_{1,1}&h_{1,2}\\
h_{2,1}&h_{2,2}
\end{pmatrix}\bm{\psi}
=\left(\frac{\omega}{c}\right)^2
\begin{pmatrix}
s_{1,1}&s_{1,2}\\
s_{2,1}&s_{2,2}
\end{pmatrix}\bm{\psi}
\end{equation}
with
\begin{equation}
h_{I,J}=\langle\bm{\nabla}\times\bm{\phi}_{\mathrm{I}}|\mu^{-1}(\bm{r})|\bm{\nabla}\times\bm{\phi}_{\mathrm{J}}\rangle,
\end{equation}
\begin{equation}
s_{I,J}=\langle\bm{\phi}_{\mathrm{I}}|\varepsilon(\bm{r})|\bm{\phi}_{\mathrm{J}}\rangle,
\end{equation}
\begin{equation}
\bm{\psi}=\begin{pmatrix}\bm{c}_1,&\bm{c}_2\end{pmatrix}^T,
\end{equation}
and $I$, $J$=1, 2.
Due to the overlap between the basis, the off-diagonal elements $h_{1,2}$, $h_{2,1}$, $s_{1,2}$, and $s_{2,1}$ are finite in general. 
Therefore, the sign of $\varepsilon$ and $\mu$ cannot be factored out, which results in the significant difference of the band structures [see Fig.~3(a)~and Fig.~6].

\begin{figure}[]
   \includegraphics[width=\hsize]{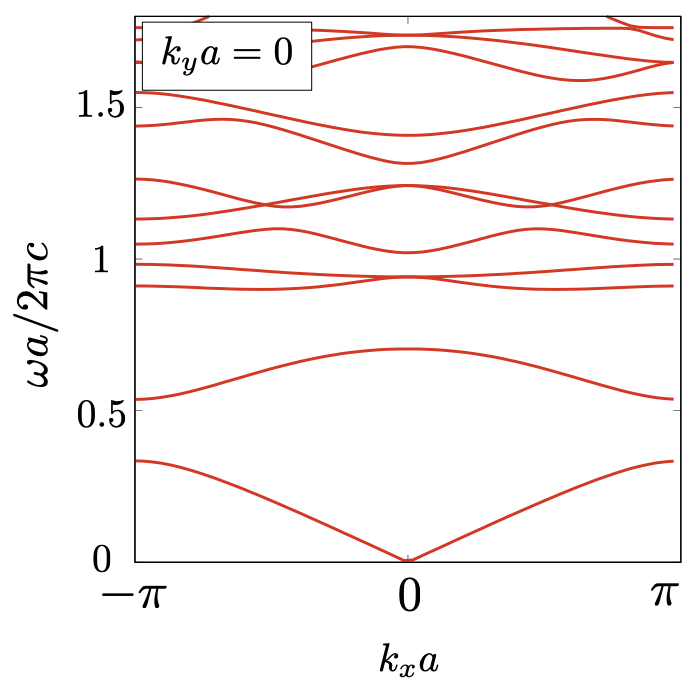}
 \caption{
 Photonic band structure of the square lattice photonic crystal with $(\varepsilon, \mu)=(5.9,0.4)$ at $k_ya=0$.
The radius of internal structures of the photonic crystal is chosen as $0.2a$ with unit-cell size $a$.
The frequency $\omega$ is real in the entire region, and EPs do not emerge.
} 
\end{figure}

\section*{APPENDIX D: Analysis of a $\omega$-dependent toy model by two different approaches}
In the main text, we have discussed the complex band structures emerging in the photonic crystal composed of NIM by considering the vertical axis of the band structure as the unit-cell size.
On the other hand, we can obtain the band structure of the frequency $\omega$ by solving the $\omega$-dependent system self-consistently.
We can see the consistency between these two different approaches by analyzing a $\omega$-dependent toy model.

Here, let us consider the following $\omega$-dependent toy model in one dimension.
\begin{equation}
\begin{pmatrix}
M_{\mathrm{L}}&\mathrm{sin}(ka)\\
\mathrm{sin}(ka)&-M_{\mathrm{L}}
\end{pmatrix}
\psi=(\omega a)^2
\begin{pmatrix}
1+M_{\mathrm{R}}(\omega)&0\\
0&1-M_{\mathrm{R}}(\omega)
\end{pmatrix}
\psi,
\end{equation}
with 
\begin{equation}
M_{\mathrm{R}}(\omega)=1-\frac{1}{\omega^2},
\end{equation}
\begin{equation}
M_{\mathrm{L}}=-0.3.
\end{equation}

We compute the band structure of $\omega$ and ``band structure" of the unit-cell size $a$. For the computation of the former bands, we solve the $\omega$-dependent model [Eq.~(23)] self-consistently by fixing $a$ to $a_0=1$.
For the computation of the latter bands, we compute eigenvalues by fixing $\omega$ to $\omega_{\mathrm{c}}$ satisfying $M_{\mathrm{R}}=1-1/\omega^2$ for given $M_{\mathrm{R}}$.
The band structure of $\omega$ is plotted in Fig.~7(a) (see the green line). In this figure, the ``band structure" of $a$ is also plotted for $M_\mathrm{R}(\omega_\mathrm{c})=-1.2$ (see the blue and red lines). The black line denotes $\omega_\mathrm{c}a_0$. This figure indicates that these eigenvalues coincide with each other (i.e., the green and blue lines cross).
We note that the imaginary part is zero at the point where the two bands cross below EPs.
For $M_{\mathrm{R}}(\omega_\mathrm{c})=-1.43$, two bands cross on the EPs [see Fig.~7(b)].
For  $M_{\mathrm{R}}(\omega_\mathrm{c})=-1.7$, two bands cross above the EPs [see Fig.~7(c)].

We stress that the band structure of $\omega$ and the ``band structure" of $a$ are essentially different.
Therefore, in Fig.~7(b), the EPs on the blue band do not emerge on the green band as EPs, despite the eigenvalues coincide with each other.
For $M_\mathrm{R}(\omega_{\mathrm{c}})=-2.1$, the two bands no longer cross.

With the above results, we can see the consistency between the band structure of $\omega$ and the ``band structure" of $a$.
The above numerical data indicate that in the region where the ``band structure" of $a$ become complex, the bands of $a$ and the band of $\omega$ do not cross the latter of which denotes a propagating mode (see green lines in Fig.~7).
With this result, we consider that the eigenmodes cannot propagate to the bulk when the bands of $a$ become complex.

\begin{figure}[]
   \includegraphics[width=\hsize]{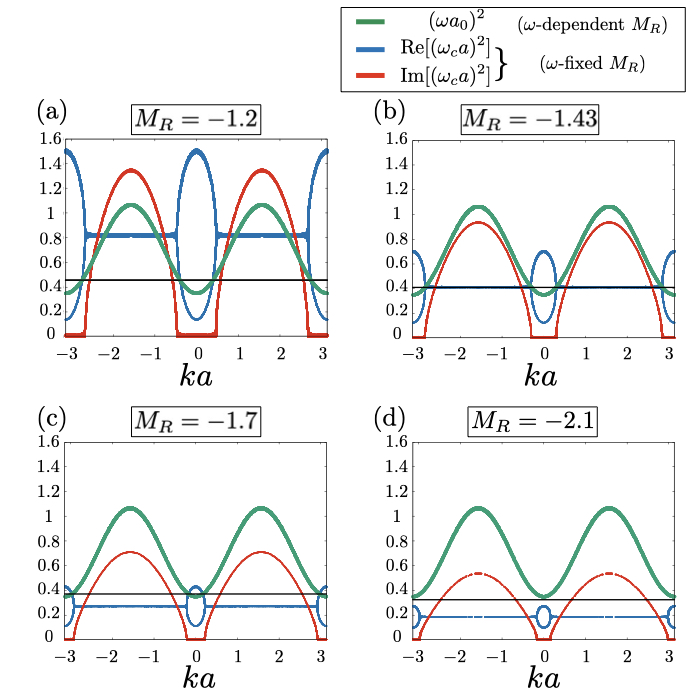}
 \caption{
Plot of the eigenvalues of Eq.~(23).
The self-consistent solution $\omega a_0$ ($a_0=1$) is plotted in green. 
The real (imaginary) part of the eigenvalues for fixed $\omega$ is plotted in blue (red).
The black line indicates $(\omega_\mathrm{c} a_0)^2$ with $\omega_{\mathrm{c}}$ for given $M_\mathrm{R}$.
Panel (a) is the plot of band structures when $M_R=-1.2$.
The green band and the blue band overlap below EPs.
Panel (b) is the plot of band structures when $M_R=-1.43$.
The green band and the blue band overlap on EPs.
Panel (c) is the plot of band structures when $M_R=-1.7$.
The green band and the blue band overlap above EPs.
Panel (d) is the plot of band structures when $M_R=-2.1$.
The green band and the blue band do not overlap.
}
\end{figure}

\end{document}